\documentclass{article}
\usepackage{graphics}
 
\def\hmpc{\rm \,h^{-1}\,Mpc}

\def\xis{$\xi(s)$\ }

\def\n_med{{\left<n\right>}}

\def\begc{\begin{center} }
\def\endc{\end{center} } 
\def\begf{\begin{figure} }
\def\endf{\end{figure} }

\def\j3{{J_3}}

\title{THE REFLEX CLUSTER SURVEY: OBSERVING STRATEGY AND 
FIRST RESULTS ON LARGE--SCALE STRUCTURE}

\author{L. Guzzo$^1$, H. B\"ohringer$^2$, P. Schuecker$^2$, 
C.A. Collins$^3$, \\
S. Schindler$^3$, D.M. Neumann$^4$, S. De Grandi$^1$, R. Cruddace$^5$, \\ 
G. Chincarini$^1,6$, A.C. Edge$^7$, P.A. Shaver$^8$, W. Voges$^2$
\\ \\ $^1$ Osservatorio Astronomico di Brera, Milano/Merate, Italy\\
$^2$ Max-Planck-Institut f\"ur extraterr. Physik, Garching, Germany\\
$^3$ Liverpool John Moores Univ., Liverpool, U.K.\\
$^4$ CEA Saclay, Service d'Astrophysique, Gif-sur-Yvette, France\\
$^5$ Naval Research Laboratory, Washington D.C., U.S.A.\\
$^6$ Dipartimento di Fisica, Universit\`a degli Studi di Milano, Italy\\
$^7$ Physics Department, University of Durham, U.K.\\
$^8$ European Southern Observatory, Garching, Germany
}

\date{}

\begin{document}

\maketitle


\section{Introduction}

As a modern version of ancient cartographers, during the last 20 years 
cosmologists have been
able to construct more and more detailed maps of the large--scale 
structure of the Universe, as delineated by the
distribution of galaxies in space.
This has been possible through the development of redshift surveys,
whose efficiency in covering ever larger volumes has increased 
exponentially thanks to the parallel evolution in the performances
of spectrographs and detectors (see e.g. Da Costa 1998 and Chincarini 
\& Guzzo 1998, for recent reviews of the historical development 
of this field).

While the most recent projects, as the Las Campanas Redshift Survey
(LCRS, Shectman et al. 1996) and the ESO Slice Project (ESP, Vettolani
et al. 1997) have considerably enlarged our view by collecting several
thousands of redshifts out to a depth of $\sim 500 \hmpc$ 
\footnote{Here h is the
Hubble constant in units of 100 km s$^{-1}$ Mpc$^{-1}$}, the quest 
for mapping a ``fair sample'' of the Universe is not yet fully over.  
These modern galaxy redshift surveys have indeed been able to show 
for the first time that large--scale structures such as surperclusters 
and voids keep sizes which are smaller than those of the surveys 
themselves (i.e. $\sim 100-200\hmpc$).  This is in contrast to the 
situation only a few years 
ago, when every new redshift survey used to discover newer and larger
structures (a famous case was the ``Great Wall'', spanning the whole
angular extension of the CfA2 survey, see Geller \& Huchra 1989).
On the other hand, it has also been realised from the ESP and LCRS 
results, among others, that to characterise statistically the scales where the 
Universe is still showing some level of inhomogeneity, surveys 
covering much larger areas to a similar depth ($z\sim 0.2$), are necessary.

To fulfill this need on one side two very ambitious galaxy survey 
projects have started\footnote{An earlier pioneering attempt in this
direction was the Muenster Redshift Project, that used 
objective--prism spectra to collect low--resolution redshifts for
a few hundred thousands galaxies (see e.g. Schuecker et al. 1996)}, 
the Sloan Digital Sky Survey (SDSS, Margon 1998) 
and the 2dF Redshift Survey (Colless 1999).  The SDSS 
in particular, including both a photometric survey in five filters and a 
redshift survey of a million galaxies over the whole North Galactic cap, 
represents a massive effort involving 
the construction of a dedicated telescope with specialised camera and 
spectrograph.

\begf
\resizebox{\hsize}{!}{\includegraphics{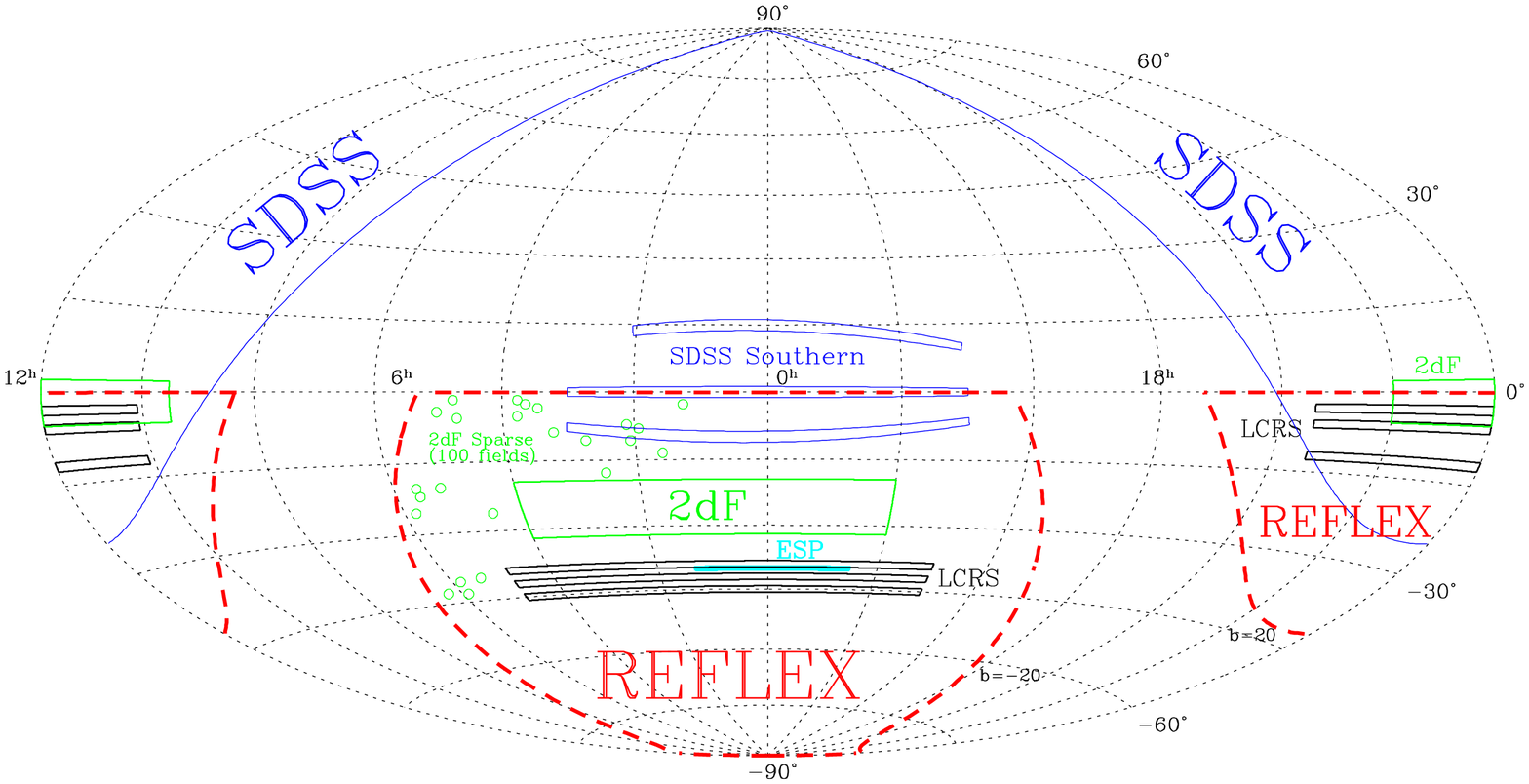}}
\caption{Aitoff view of how wide--angle redshift surveys 
with typical depth of $z\sim 0.2$ are covering the available sky 
(see Guzzo 1999 for details).}
\label{fig:surveys}
\endf
Alternatively, one could try to cover similar volumes using 
a different tracer of large--scale structure, 
one requiring a more modest investment in telescope time, so that a survey
can be performed using standard instrumentation and within the 
restrictions of public telescopes as those of ESO.  
This has been the strategy of the REFLEX Key Programme, 
that uses {\bf clusters of galaxies} to cover, in the Southern 
hemisphere, a volume of the Universe comparable to that of the 
SDSS in the North.  Clusters, being rarer than galaxies,
are evidently more efficient tracers for mapping very
large volumes.  The price to pay is obviously that of loosing 
resolution in the description of the small--scale 
details of large--scale structure.  However such information is 
already provided by the present generation of galaxy surveys.

The REFLEX cluster survey, in particular, is based on clusters 
selected through their X--ray
emission, which is a more direct probe of their mass content
than simple counts of galaxies, i.e. richness.
Therefore, using clusters as
tracers of large--scale structure, not only do we sample very
large scales in an observationally efficient way, but if 
we select them through their X--ray emission we also have 
a more direct relation between luminosity and mass.  
Further benefits of the extra information provided by the
X-ray emission have been
discussed in our previous Messenger article (B\"ohringer et al. 
1998).  In the same paper, we also presented details on
the cluster selection, the procedure for measuring of X-ray fluxes, 
and some properties of the objects in the catalogue, as the
X-ray luminosity function, together with a first determination
of the power spectrum.  Results on the cluster X-ray luminosity 
function have also been obtained during the
development of the project from an early pilot subsample at bright fluxes 
(see De Grandi et al. 1999).
Here we would like to concentrate on the key features of the optical 
follow-up observations at ESO, and then present more results on the
large-scale distribution of REFLEX clusters and their clustering
properties.  Nevertheless, for the sake of clarity we shall first 
briefly summarise the general features of the survey.

\section{The REFLEX Cluster Survey}

The REFLEX (ROSAT-ESO 
Flux Limited X-ray) cluster survey combines the X--ray data from the ROSAT 
All Sky Survey (RASS), and ESO optical observations to construct a complete 
flux--limited sample of about 700 clusters with measured redshifts and 
X-ray luminosities.   The survey covers almost the southern celestial 
hemisphere (precisely $\delta<2.5^\circ$), at galactic latitude 
$|b_{II}|>20^\circ$ to avoid high NH column densities and crowding by stars. 
Figure~\ref{fig:surveys} 
(reproduced from Guzzo 1999), shows the region covered by the 
REFLEX survey, together with those of wide--angle galaxy redshift surveys of
similar depth, either recently completed (ESP, LCRS), or just started 
(SDSS, 2dF).  

\begf
\resizebox{\hsize}{!}{\includegraphics{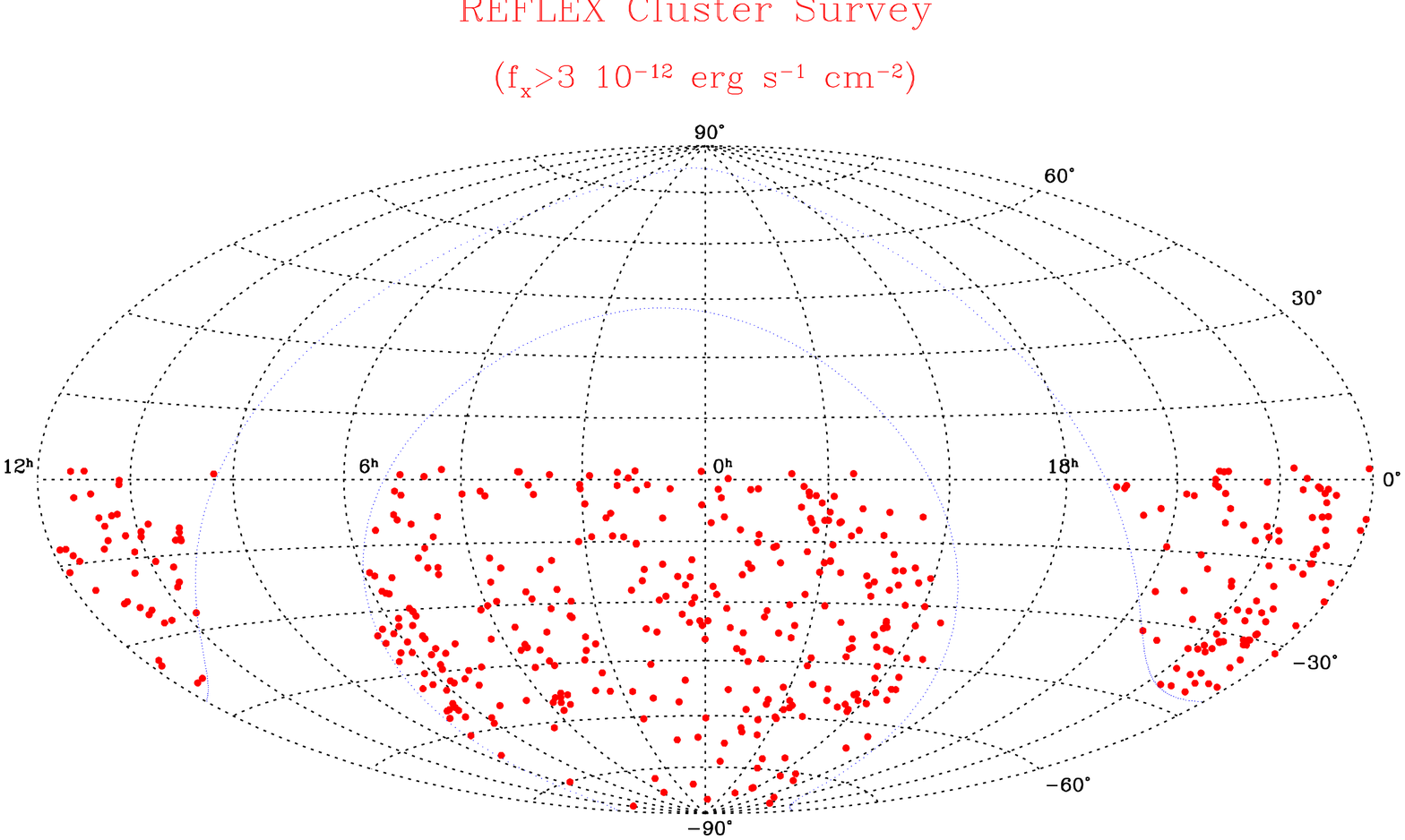}}
\caption{The distribution on the sky of the clusters in the REFLEX sample
with $f_x> 3 \times 10^{-12}$ erg s$^{-1}$ cm$^{-2}$.
Note that the area around the Magellanic Clouds 
($\alpha\sim1^h$ and $\sim5^h$, $\delta\sim -70^\circ$), is 
not included in the survey.} \label{fig:aitoff}
\endf
Presently, we have selected a complete sample 
of 460 clusters to a nominal flux limit of $3 \times 10^{-12}$ erg 
s$^{-1}$ cm$^{-2}$ (in the ROSAT band, 0.1--2.4 keV).  
Due to to the varying RASS exposure time and  
interstellar absorption over the sky, when a simultaneous requirement 
of a minimum number of source counts is applied, some parts
of the sky may reach a lower flux.  For example, when computing the
preliminary REFLEX luminosity function presented in B\"ohringer et al 
(1998), a threshold of 30 source counts in the ROSAT hard band was set,
which resulted in a slightly reduced flux limit over 21.5\% of the survey 
area.  The important point to be made is that a full map of these 
variations is known as a function of angular position and is 
exactly accounted for in all the statistical analyses.

This first sample, upon which the discussion on clustering and large-scale 
structure presented here is based, has been constructed to be at least 90\% 
complete, 
as described in B\"ohringer et al. (1998). Several external checks, as 
comparisons with independently extracted sets of clusters, support
this figure.  The careful reader may have noticed that the 
total number of objects in this sample has been reduced by 15 since 
the time of writing our previous Messenger report.  
This is the result of the ongoing process of final identification and 
redshift measurement: spectroscopy and detailed deblending of X-ray 
sources have led to a reclassification of 15 X-ray sources which either 
had AGN counterparts or fell below the flux limit.
With these new numbers, as of today 95\% of the 460
candidates in this sample are confirmed and observed spectroscopically.
Final clearing up and measurement of redshifts for the remaining 
$\sim 20$ candidates is foreseen for a forthcoming observing run 
next May.   In the following, we shall simply refer to this complete sample
as the ``REFLEX sample''.  The distribution on the sky of the REFLEX
clusters defined in this way is shown in Figure~\ref{fig:aitoff}.

\section{Optical Follow-up Observing Strategy}

The follow--up optical observations of REFLEX clusters were 
started at ESO in 1992, under the status of a Key Programme.
The goal of these observations was twofold: a) obtain a definitive 
identification of ambiguous candidates; b) obtain a measurement of 
the mean cluster redshift.  

First, a number of candidate clusters 
required direct CCD imaging and/or spectroscopy to be safely included 
in the sample.  
\begf
\begin{center}
\resizebox{\hsize}{!}{\includegraphics{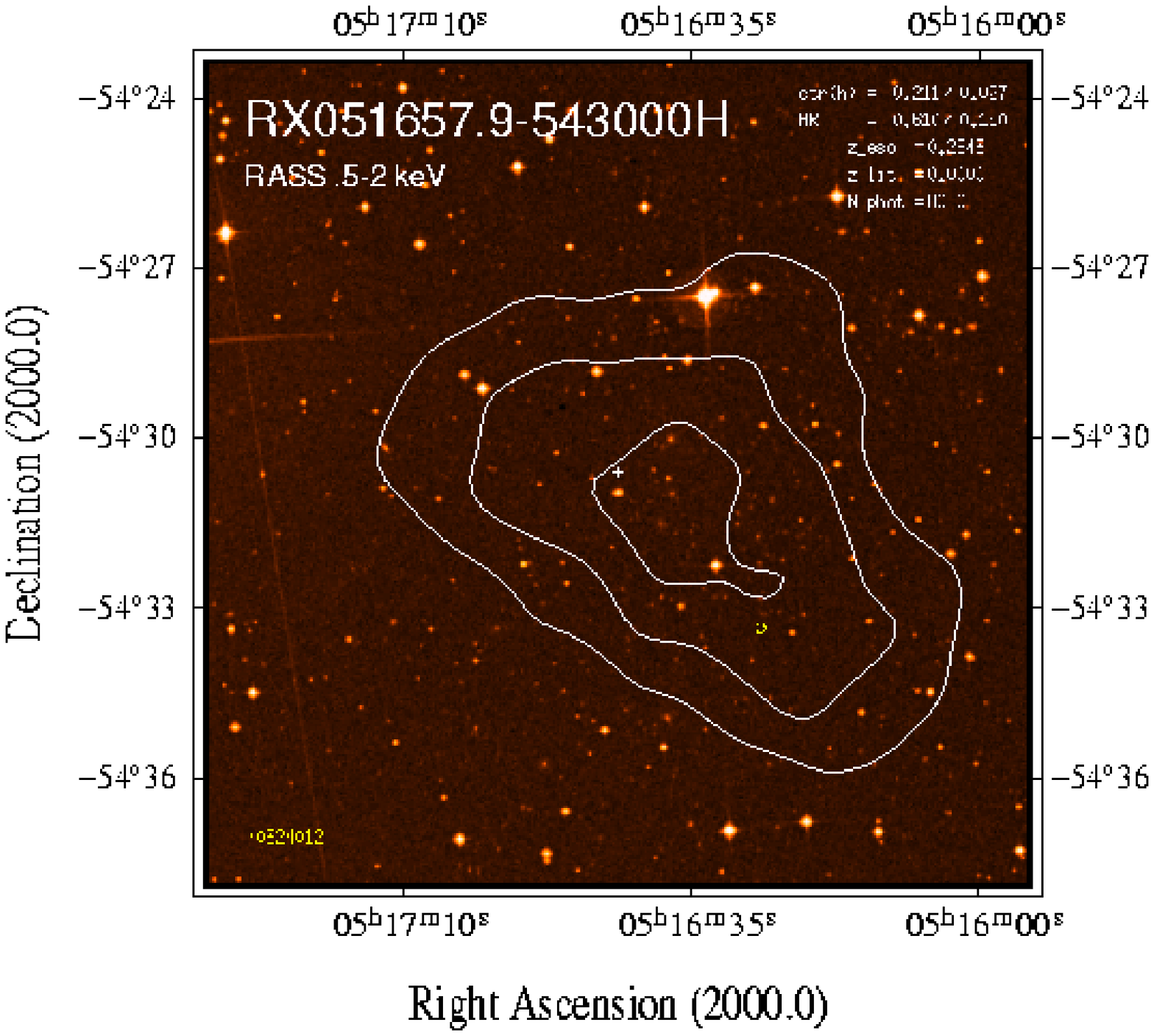}}
\end{center}
\caption{$15\times 15$ arcmin identification image of a REFLEX cluster 
at z=0.294. The optical image is from the Digitised Sky Survey (DSS), 
with superimposed the X--ray contours from the RASS.  As suggested by 
the contours, this X-ray source shows a statistically significant extention.}
\label{fig:clus_dss}
\endf
For example, candidates characterised by a poor appearance on the Sky Survey
IIIa-J plates, with no dominant central galaxy or featuring a point--like 
X--ray emission had to pass further investigation.  In this case, 
either the object at the X-ray peak was studied 
spectroscopically, or a short CCD image plus a spectrum
of the 2--3 objects nearest to the peak of the X-ray 
emission was taken. This operation was preferentially 
scheduled for the two smaller telescopes (1.5~m and 2.2~m, see below),
and was necessary to be fully 
sure of keeping the completeness of the selected sample close to
the desired value of 90\%.   In this way, a number of AGN's were 
discovered and rejected from the main list.

Once a cluster was identified, the main scope of the optical observations 
was then to secure a reliable redshift.   The observing 
strategy was designed so as to compromise between the desire of having
several redshifts per cluster, coping with the multiplexing limits of
the available instrumentation, and the large number of clusters to be
measured.   Previous 
experience on a similar survey of EDCC clusters (Collins et al. 
1995), had shown the importance of not relying on just one or two galaxies 
to measure the cluster redshift, especially for clusters without a dominant 
cD galaxy.   EFOSC1 in MOS mode was a perfect instrument for 
getting quick redshift measurements for 10--15 galaxies at once, but only 
for systems that could reasonably fit within the small field of view of 
the instrument (5.2 arcmin in imaging with the Tektronics CCD \#26,
but less than 3 arcmin for spectroscopy in MOS mode, 
due to hardware/software limitations in the use of the MOS masks).   
This feature clearly made this combination useful only for 
clusters above z=0.1, i.e. where at least the core region could be 
accommodated within the available area.   

The other important aspect of using such an instrumental set--up is that 
in several cases, after removal of background/foreground objects one is 
still left with 8-10 galaxy redshifts within the cluster, by which a 
first estimate of the cluster velocity dispersion can be attempted.  
This clearly represents further, extremely important information 
related to the cluster mass, especially when coupled to the X-ray
luminosity available for all these objects.    

\begf
\begin{center}
\resizebox{\hsize}{!}{\includegraphics{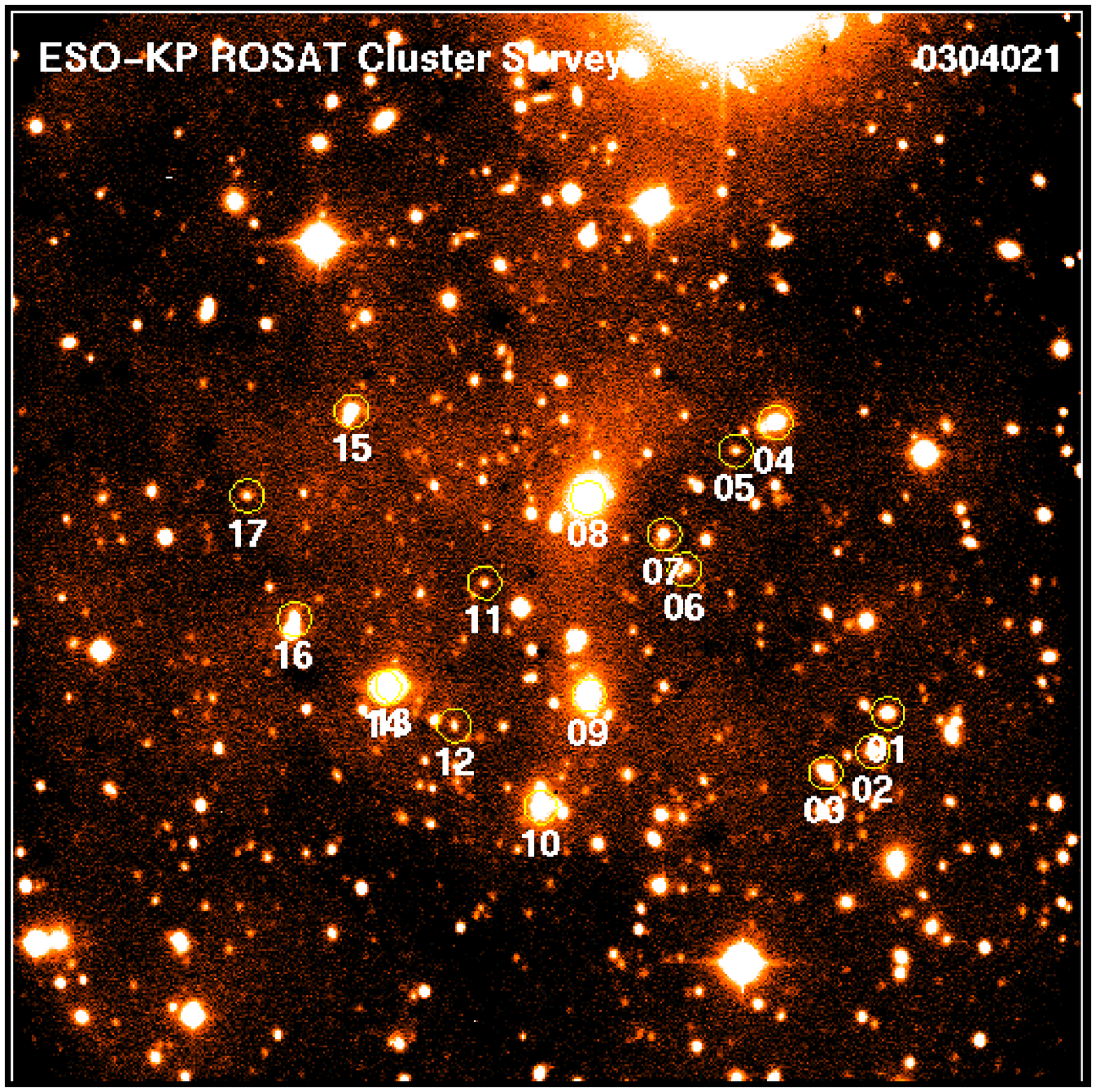}}
\end{center}
\caption{Short CCD exposure ($\sim 60$ sec) taken in white light 
with EFOSC1 at the 3.6~m telescope, covering the 
central region of the REFLEX cluster of previous figure.  The field
side is 5.2 arcmin.  Note the remarkable glory of faint galaxies around 
the central brightest members.  The numbers indicate the positions of 
the slitlets for the MOS observations.}
\label{fig:clus_ccd}
\endf

At smaller redshifts, doing efficient multi--object spectroscopy work would 
have required a MOS spectrograph with a larger field of view, i.e. 20--30 
arcminutes diameter.  One possible choice could have been the fibre 
spectrograph 
Optopus (Avila et al. 1989), but its efficiency in terms of numbers of 
targets observable 
per night was too low for covering several hundred 
clusters as we had in our sample.  The best solution in terms of telescope
allocation pressure and performances was found in using single--slit 
spectroscopy and splitting the work between the 1.5~m and 2.2~m telescopes.
Clearly, this required accepting some compromise in our wishes of
having multiple redshifts, so that at the time of writing about 30\%
of the clusters observed at ESO have a measure based on less than 3 
redshifts.  The reliability of these as estimators of the cluster 
systemic velocity, 
however, is significantly enhanced by the coupling of the galaxy positions 
with the X--ray contours: we can clearly evaluate which galaxies have the
highest probability to be cluster members.  This is another advantage 
of a survey of X--ray clusters over optically--selected clusters.
At the time of writing, in about seven years of work,  we have observed 
spectroscopically a total of about 500 cluster candidates, collecting over 
3200 galaxy spectra.

Figure~\ref{fig:clus_dss} shows one example of the identification images,
used for the first confirmation of the candidates.  These
pictures are constructed for all our candidates by combining the Digitised 
Sky Survey plates of the ESO/SRC atlas (image) and the RASS X--ray data 
(contours).  Although this cluster (RX051657.9-543000), 
lies at a redshift z=0.294 (close to the redshift identification limit of 
REFLEX at z=0.32), a sufficient number of galaxies is detected 
in the optical image even at the depth of the survey plates.  This
picture gives a clear example of how the X--ray contours ``guide the eye''
in showing which are the ``best'' galaxies to be observed.  In this
case we had a MOS observation, but had we observed only the central
two galaxies within the X--ray peak, we would have obtained a correct 
estimate of the cluster redshift.

The same cluster is then visible in the direct EFOSC1 image of  
Figure~\ref{fig:clus_ccd}.  This is a short service exposure 
(less than one minute),
taken in white light as a template for the drilling of the slits on the EFOSC1
MOS mask.   The image shows a spectacular abundance of faint galaxies,
one of the most impressive cases observed during our survey.

\begf
\resizebox{\hsize}{!}{\includegraphics{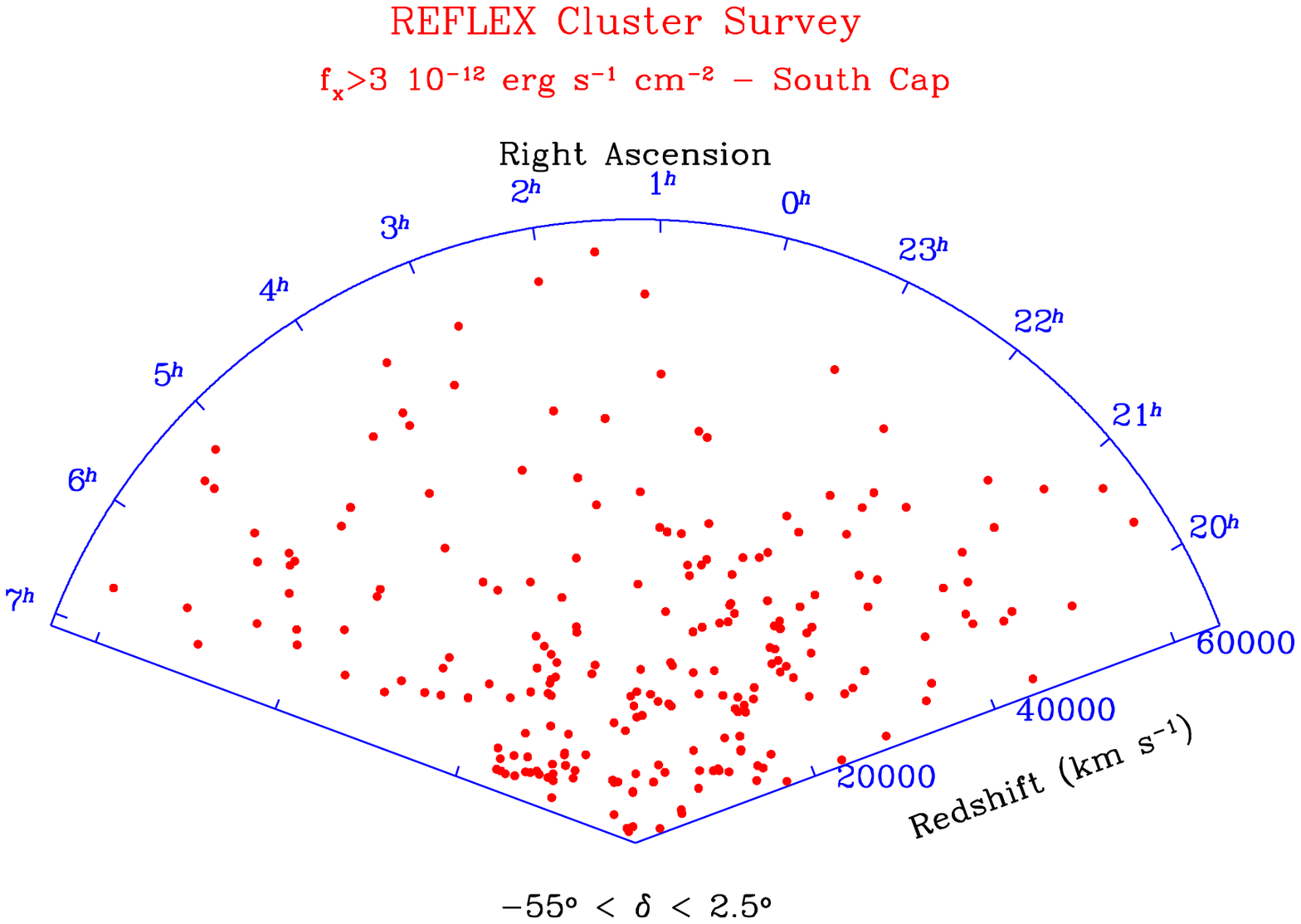}}
\caption{The large--scale distribution of REFLEX clusters within part of
the South Galactic Cap region.  Only objects North of $\delta=-55$ are shown,
to avoid excessive confusion by the projection along declination.}
\label{fig:coneplot}
\endf

After looking at this latest picture, it is natural to ask what is, 
for example, the galaxy luminosity function of this cluster, or what
are the colours of this large population of faint objects.  This is
clearly important information, which at the moment, however, is only available 
for a restricted fraction of REFLEX clusters (Molinari et al. 1998).  
To cover this aspect, a wide--field imaging campaign is going to commence 
in the next semester, starting first with those
medium--redshift clusters that best match the WFI at the 
2.2~m telescope.

\section{The Large--Scale Distribution of X--Ray Clusters}

The cone diagram of Figure~\ref{fig:coneplot} plots the distribution 
of REFLEX clusters in the South Galactic cap
area of the survey, selecting only objects with $z<0.2$ and 
$\delta>-55^\circ$, to ease visualization.  
One can easily notice a number of superstructures with sizes $\sim 100 
\hmpc$, that show explicitly the typical scales on which  
the cluster distribution is still inhomogeneous.

\begf
\begin{center}
\resizebox{10cm}{!}{\includegraphics{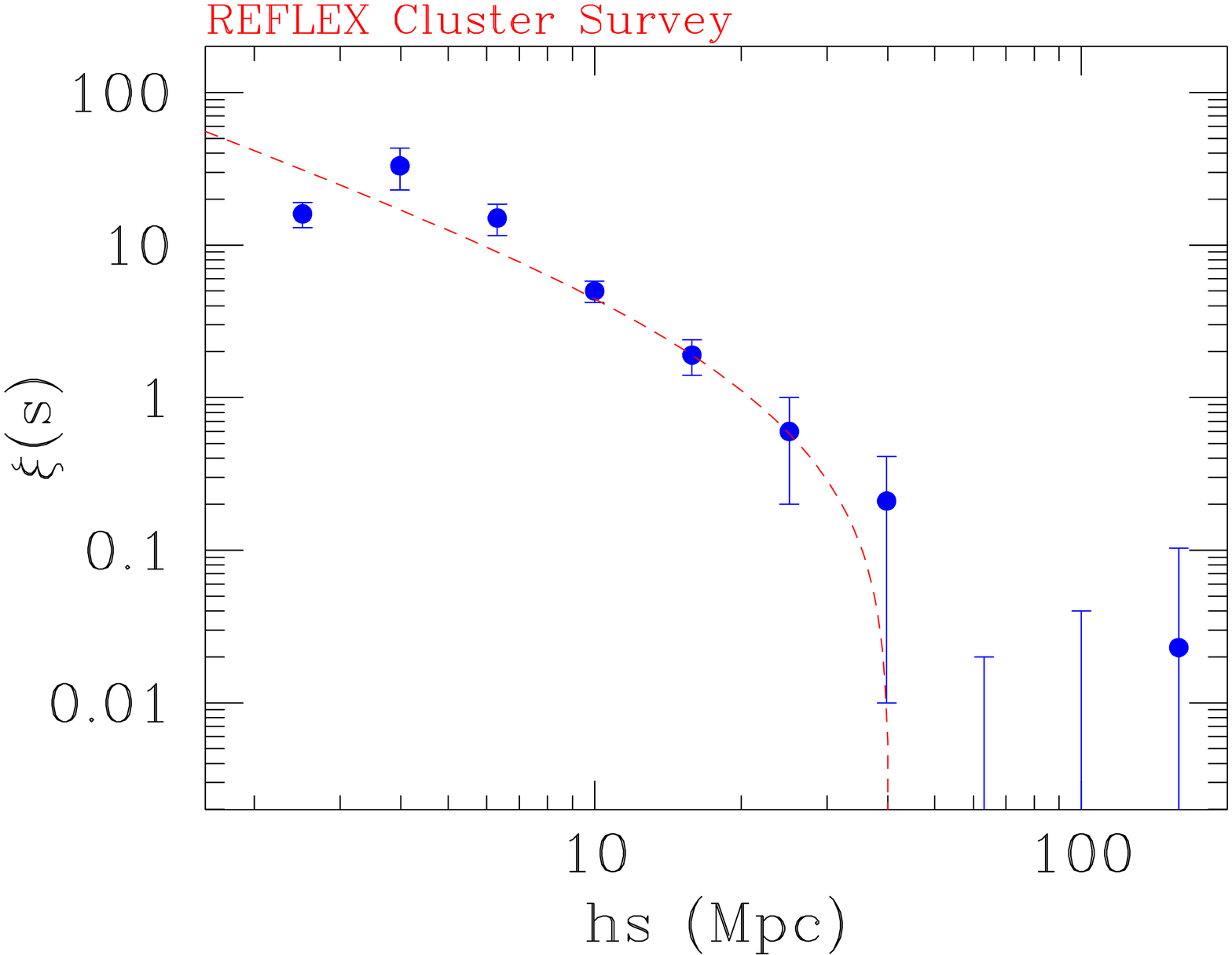}}
\end{center}
\caption{The two--point correlation function of clusters of galaxies
in the REFLEX survey.  The filled circles give the direct estimate,
while the dashed line is computed by Fourier transforming the power 
spectrum of the survey data, that we showed in B\"ohringer 1998.}
\label{fig:xi}
\endf
This inhomogeneity can be quantified at the simplest level through the
two--point correlation function \xis, that measures the probability in
excess of random of finding a pair of clusters with a given separation,
(the variable $s$ is used here to indicate separations in {\it redshift
space}).  A preliminary estimate of \xis for the REFLEX sample is 
shown by the filled circles in Figure~\ref{fig:xi}.  The 
dashed line shows for comparison the Fourier transform of a 
simple fit to the power spectrum P(k),
measured from a subsample of the same data (see B\"ohringer et al. 1998).
The estimates of \xis and P(k) were performed 
by taking into account the angular dependence of the survey
sensitivity, i.e., the exposure map of the ROSAT All-Sky Survey, and
the NH map of galactic absorption.  
The good agreement between the two curves is on one side an indication of the 
self--consistency of the two estimators applied in redshift and Fourier space.
Also, it shows a remarkable stability in the clustering properties of
the sample, given that for the measure of P(k) the data were conservatively
truncated at a comoving distance of $200\hmpc$ (Schuecker et al., 
in preparation), while the estimate of \xis uses the whole catalogue 
(Collins et al., in preparation).  On the other hand, this also indicates 
that the bulk of the clustering signal on \xis is produced within the inner 
$200\hmpc$ of the survey, which is expected because this is the most densely 
sampled part of the flux--limited sample.  We shall explore this in more 
detail when studying volume--limited samples, with a well--defined lower 
threshold in X-ray luminosity.

Figure~\ref{fig:xi} shows that \xis for clusters of galaxies is fairly well
described by a power law out to $~40\hmpc$, and
then breaks down, crossing the zero value around $50\hmpc$.  It is interesting
to compare it with the two--point correlation function of
galaxies, as we do in Figure~\ref{fig:xi_gal}.  The galaxy data 
shown here (points), are obtained from two volume--limited subsamples 
of the ESP survey (Guzzo et al. 1999).   They are compared to \xis
from the REFLEX clusters, given by the dashed lines.  To ease the 
comparison, we preferred to plot the curves of \xis as computed from
the Fourier transform of P(k).  This is given by the top line,
while the bottom line has been re--scaled by a factor $b_c^2=(3.3)^2$
in amplitude.
This difference in amplitude, or {\it bias}, is expected, as clusters 
represent the high, rare peaks of the galaxy density distribution,
and it can be demonstrated (Kaiser 1984), that their clustering has to
be enhanced with respect to that of the general field.  This quoted
value of $b_c$, however, does not have a direct physical interpretation, 
as it is obtained from a flux--limited sample, and thus related 
to clusters having different mean intrinsic luminosities at different 
distances.  One important aspect of selecting clusters through their X--ray
emission is in fact that a selection in X-ray luminosity is closer to
a selection in {\it mass}, than if one used a measure like the cluster
{\it richness} (i.e. the number of galaxies within a given radius and
a given magnitude range, as in the case of the Abell catalogue).  
For this reason, the observation that \xis has a different
amplitude for volume--limited subsamples with different X--ray 
luminosity limits (i.e. a different value of $b_c$), has important
implications for the theory (Mo \& White 1996).
The validity of a simple biasing 
amplification mechanism on these scales is explicitly supported by the very 
similar slopes of \xis shown in Figure~\ref{fig:xi_gal} by galaxies and 
clusters.  A full discussion concerning the luminosity dependence of 
the amplitude of the correlation function is being prepared (Collins et al.,
in preparation).
\begf
\begin{center}
\resizebox{10cm}{!}{\includegraphics{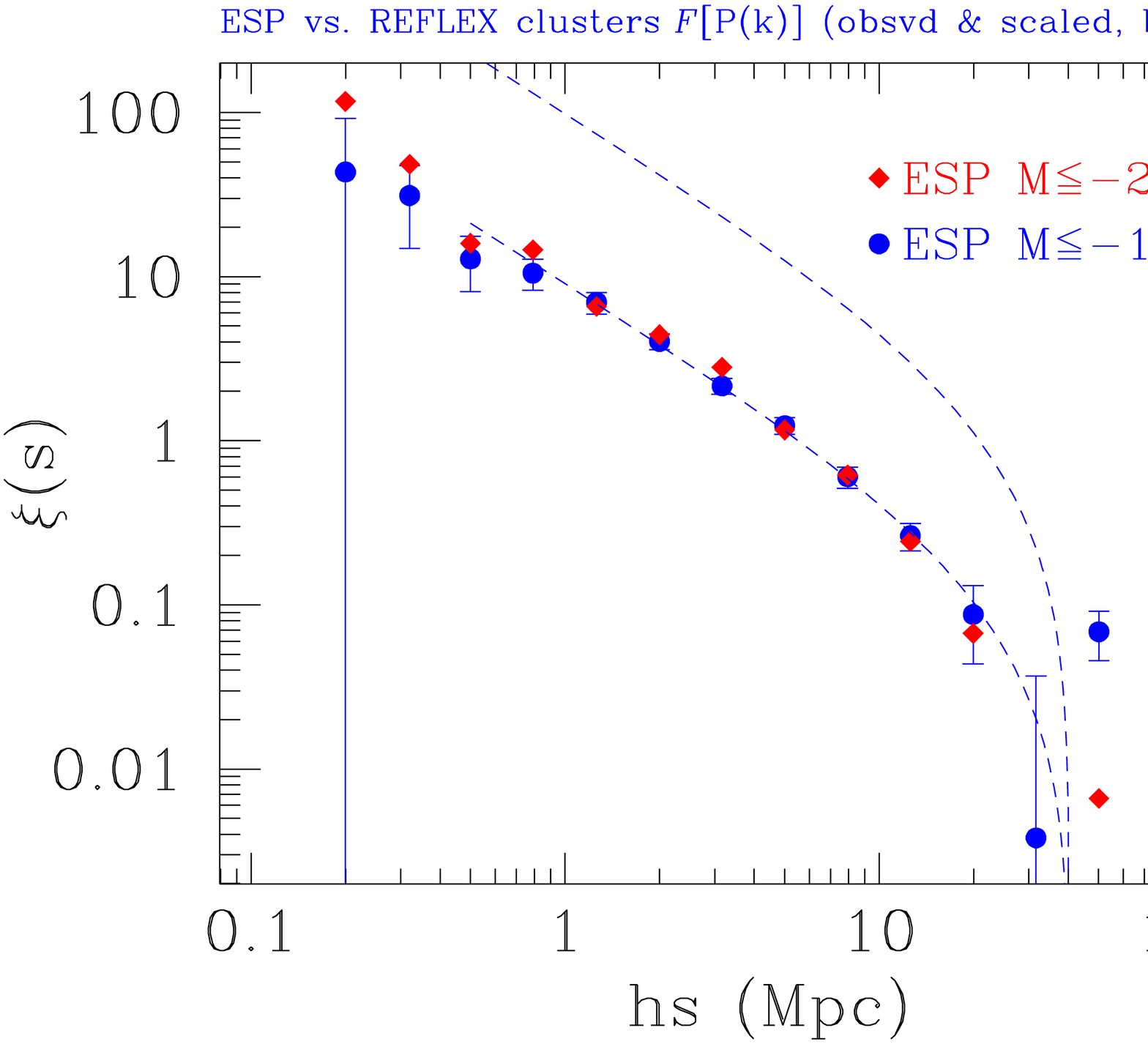}}
\end{center}
\caption{Comparison of the two--point correlation function of
REFLEX clusters (top dashed line), to that of ESP galaxies (dots, 
Guzzo et al. 1999). The bottom dashed line is the REFLEX \xis after
scaling by an arbitrary bias factor of $b_c^2=(3.3)^2$.  The agreement
in shape between galaxies and clusters is remarkable. 
}
\label{fig:xi_gal}
\endf

\section{Conclusions}

We have discussed how after several years of work with the X-ray data 
from the RASS and with ESO telescopes, the 
REFLEX project has reached a stage 
where the first significant results on large--scale structure can 
be harvested.  We hope to have at least given a hint of how
the REFLEX survey is possibly the first X-ray selected cluster
survey where the highest priority was given to the statistically
homogeneous sampling over very large solid angles.  This makes it
an optimal sample for detecting and quantifying the spatial
distribution of the most massive structures in the Universe,
as can be appreciated from the high S/N superclusters visible
in the cone diagram of Figure~\ref{fig:coneplot} out to at
least $z\sim 0.1$.

We have also shown how these results have been reached
without specially designed instrumentation. Started in May 1992, 
with the next May run it will be seven years that we have
been observing and measuring redshifts for REFLEX clusters.  
Again, this long timescale is a consequence of the fact that the 
project is based on ``public'' instrumentation, and therefore
subject to the share of telescope time with the general ESO community.
Clearly, even in these terms, this could have never been possible at 
ESO without the long--term ``Key Programme'' scheme.  In
fact it is exactly for surveys like REFLEX, 
i.e. large cosmology projects, that the concept of Key Programmes
was originally conceived, under suggestion of those few people that 
at the end of the eighties were starting doing large--scale structure 
work in continental Europe.  If the aim of the Key Programmes was 
that of making the exploration of the large-scale structure of the
Universe feasible for European astronomers, providing them with a 
way to compete with the dedicated instrumentation of other institutions,
we can certainly say that, after ten years, this major goal has probably 
been reached.

We thank Harvey MacGillivray and Paolo Vettolani for their
support to the birth of this project and all the people who
have contributed in different ways to its realization.

%


\begin{thebibliography}{}

\bibitem[]{} Avila, G., D'Odorico, S., Tarenghi, M., \& Guzzo, L., 1989,
The Messenger, 55, 62
\bibitem[]{} B\"ohringer, H., et al., 1998, The Messenger, 94, 21
(astro-ph/9809382)
\bibitem[]{} Chincarini, G., \& Guzzo, L., 1998, in {\it Proc. of
V-eme Colloque de Cosmologie}, H. De Vega ed., in press
\bibitem[]{} Colless, M., 1998, Phil. Trans. R. Soc.
Lond. A, in press (astro-ph/9804079)
\bibitem[]{} Collins. C.A., Guzzo, L., Nichol, R.C., \& Lumsden,
S.L., 1995, MNRAS, 274, 1071
\bibitem[]{} Da Costa, L.N., 1998, in {\it Evolution of Large--Scale
Structure: from Recombination to Garching}, T. Banday \& R.
Sheth eds., in press
(astro-ph/9812258)
\bibitem[]{} De Grandi, S., et al., 1999, ApJ (Letters), in press
(astro-ph/9812423)
\bibitem[]{} Geller, M.J. \& Huchra, J.P., 1989, Science, 246, 897
\bibitem[]{} Guzzo, L., 1999, in {\it Proc. of XIX Texas
Symposium}, (Paris - October 1998), Elsevier, in press
\bibitem[]{} Guzzo, L., et al. (The ESP Team), 1999, A\&A,
in press (astro-ph/9901378)
\bibitem[]{} Kaiser, N., 1984, ApJ, 284, L9
\bibitem[]{} Margon, B., 1998, Phil. Trans. R. Soc. Lond. A,
in press (astro-ph/9805314)
\bibitem[]{} Mo, H.J., \& White, S.D.M., 1996, MNRAS, 282, 347
\bibitem[]{} Molinari, E., Moretti, A., Chincarini, G., \& De Grandi,
S., 1998, A\&A, 338, 874
\bibitem[]{} Schuecker, P., Ott, H.-A., \& Seitter, W.C., 1996, ApJ,
472, 485
\bibitem[]{} Shectman, S.A., et al., 1996, ApJ, 470, 172
\bibitem[]{} Vettolani, et al. (the ESP Team), 1998, A\&AS,
130, 323

\end{thebibliography}
\end{document}